\documentclass{article}[11pt]  
\usepackage{amsmath}
\usepackage{amssymb}
\usepackage{amsthm}
\usepackage{ifpdf}
\usepackage{wrapfig}
\usepackage{fullpage}
\usepackage{cite}
\usepackage{subcaption}
\usepackage{float}
\usepackage{enumitem}
\usepackage{graphicx}
\usepackage[noend]{algorithm2e}
\usepackage{multirow}

\usepackage{authblk}

\theoremstyle{definition}
\newtheorem{theorem}{Theorem}[section]
\newtheorem{lemma}[theorem]{Lemma}
\newtheorem{corollary}[theorem]{Corollary}

\newtheorem{definition}[theorem]{Definition}

\newtheorem{observation}[theorem]{Observation}

\newcommand{\para}[1]{\vspace*{.1cm}\noindent\textbf{#1}}

\title{Reachability in Restricted Chemical Reaction Networks \footnote{This research was supported in part by National Science Foundation Grants CCF-1817602 and CCF-2329918.
}}

\author[1]{Robert M. Alaniz}
\author[1]{Bin Fu}
\author[2]{Timothy Gomez}
\author[1]{Elise Grizzell}
\author[3]{Andrew Rodriguez}
\author[2]{Marco Rodriguez}
\author[1]{Robert Schweller}
\author[1]{Tim Wylie}

\affil[1]{University of Texas Rio Grande Valley, Edinburg, TX, USA.\\ \{robert.alaniz01,bin.fu,elise.grizzell01,robert.schweller,timothy.wylie\}@utrgv.edu}
\affil[2]{Massachusetts Institute of Technology, Cambridge, MA, USA.\\\{tagomez7,marcordz@mit.edu\}@mit.edu}

\affil[3]{Texas State University, San Marcos, TX, USA.\\ andrew.rodriguez@txstate.edu}

\begin{document}

\date{}
\clearpage\maketitle
\thispagestyle{empty}

\vspace*{-.5cm}
\begin{abstract}

The popularity of molecular computation has given rise to several models of abstraction, one of the more recent ones being Chemical Reaction Networks (CRNs). These are equivalent to other popular computational models, such as Vector Addition Systems and Petri-Nets, and restricted versions are equivalent to Population Protocols. This paper continues the work on core \emph{reachability} questions related to Chemical Reaction Networks; given two configurations, can one reach the other according to the system's rules? With no restrictions, reachability was recently shown to be Ackermann-complete, which resolved a decades-old problem. 

In this work, we fully characterize monotone reachability problems based on various restrictions such as the allowed rule size, the number of rules that may create a species ($k$-source), the number of rules that may consume a species ($k$-consuming), the volume, and whether the rules have an acyclic production order (\emph{feed-forward}). We show PSPACE-completeness of reachability with only bimolecular reactions in two-source and two-consuming rules. This proves hardness of reachability in a restricted form of Population Protocols. This is accomplished using new techniques within the motion planning framework. 

We give several important results for feed-forward CRNs, where rules are single-source or single-consuming. We show that reachability is solvable in polynomial time as long as the system does not contain special \emph{void} or \emph{autogenesis} rules. We then fully characterize all systems of this type and show that with void/autogenesis rules, or more than one source and one consuming, the problems become NP-complete.
Finally, we show several interesting special cases of CRNs based on these restrictions or slight relaxations and note future significant open questions related to this taxonomy.

\end{abstract}


\section{Introduction}
The popularity of molecular computation and the need to model distributed reactions has given rise to several models of abstraction and multiple areas of research. Many of these models arose naturally in different fields decades apart, yet mathematically are nearly equivalent. The focus of this paper is on Chemical Reaction Networks (CRNs) \cite{Aris:1965:ARMA, Aris:1968:ARMA}, a model equivalent \cite{Cook:2009:AB} to Vector Addition Systems (VASs) \cite{Karp:1969:JCSS} and Petri-Nets \cite{Petri:1962:PHD}. Further, the Population Protocols model \cite{Angluin:2006:DC} is just a restricted version of these models, limited by the number of input and output elements in each operation.

Although these models may be substantively equivalent, we focus on CRNs for two reasons: first, due to the simplicity and convenience of the system definition. Specifically, expressing the production operations through reaction rules gives a straightforward rubric to measure and characterize the power of the system. The second is because CRNs, in some form, are also the oldest formulation of these types of distributed systems although their definition evolved considerably early on. 
Even though the reaction rule description is more intuitive in many cases, in our formal definitions, we rely on a matrix interpretation for notational convenience with precision, similar to Vector Addition Systems.

From the inceptions of each of these models, two of the most fundamental questions have always been \emph{production} and \emph{reachability}. Production asks if some species/element can ever be produced. Reachability simply asks if, given a system and initial configuration, whether it will even reach another specific configuration.
As the expectation of behaviors and the attainability of end goals are necessary in the design of any of these systems, production and reachability are at the heart of system design.
Despite the fundamental nature of reachability questions, most remain open or have only recently been solved, while production is better understood overall. Since many results are applicable, or equivalent, between these models, we briefly survey the reachability question in relation to each of the models.

\subsection{An Overview of Related Models}
While reachability was proven EXPSPACE-hard 1976 in \cite{Lipton:1976:TECH} and decidable in 1981 \cite{Mayr:1981:STOC}, completeness was not proven until 2021 when the unrestricted version of the reachability problem was proven to be Ackermann-complete \cite{czerwinski:2022:FOCS,leroux:2022:FOCS} for the equivalent models of Vector Addition Systems, Petri-Nets, and Standard CRNs. 

\para{Chemical Reaction Networks.} The desire and attempt to bring a mathematical abstraction to model chemical reactions has been well-documented for over a hundred years, but began in its modern incarnation during the 1960s as chemical reaction network theory \cite{Aris:1965:ARMA,Aris:1968:ARMA}. CRNs consist of a set of rules/reactions, e.g. $A+B \rightarrow C+D+E$, that represent the act of replacing the reactants ($A+B$) with the products ($C+D+E$).  
In \emph{Proper} CRNs, where reactions do not change the system's total volume, i.e., the number of reactants matches the number of products for each system rules, reachability and production are known to be PSPACE-hard even with catalytic reactions \cite{Esparza2019}. 

\para{Vector Addition Systems.} Vector Addition Systems (VAS) have an initial vector of non-negative integers and a set of operational integer vectors. The system is allowed to add these vectors however desired, as long as none of the values in the initial/counter vector ever become negative. VAS is closely related to CRNs as a reaction $A+B \rightarrow C+D+E$ can be represented by an operational vector $[ -1, -1, 1, 1, 1 ]$ assuming the vector indices map to $A,B,C,D,E$.  VAS therefore encompasses exactly the CRN models that do not make us of \emph{catalytic} reactions that utilize a species as both a reactant and a product.

The related questions of coverage, asking if there exists a reachable configuration with at least the given number of each species, and boundness, asking if the set of reachable configurations is finite, were shown to be in EXPSPACE \cite{rackoff1978covering}. Other relevant reachability results include, VASs with dimension $\leq 5$ were proven decidable \cite{hopcroft1979reachability}, as the set of reachable configurations is a semilinear set. 
In VAS with states (VASS), it was also shown that there exists $6$-dimensional VASs whose reachable configurations can form non-semilinear sets. 
In 2-Dimensional VASS, reachability is PSPACE-complete if the values of the input and target vector are encoded in binary and NL-complete if encoded in unary \cite{blondin2021reachability}.

\para{Petri-Nets.} Petri-Nets were first formally introduced in \cite{Petri:1962:PHD} in 1962 to visually describe chemical processes. Petri-Nets provide a graphical representation of reactions by having \emph{places} with edges to \emph{transitions}, which have edges to other places.  In terms of CRNs, \emph{places} correspond to the different species types, and \emph{tokens} in the Petri-Net model represent integer counts of each species type.  Weights accompanying reaction edges corresponding to consumption within reactions between different elements. Reachability in Petri-Nets has continued to be an active area of research, with numerous extensions and restrictions on the graph structure. Immediate observation Petri-Nets (IO Nets), introduced in \cite{Esparza2019}, place a restriction on the transitions such that each transition involves 3 places: the source, the destination, and the observed place. The observed place acts as a `catalyst' in the transition. By simulating a bounded-tape Turing machine, reachability in IO nets is shown to be PSPACE-hard \cite{Esparza2019}.



\para{Population Protocols.}
 Created to model distributed and decentralized computing with agents that are highly resource-limited, Population Protocols was introduced in 2004 \cite{Angluin:2004:PODC,Angluin:2006:DC}. In the model, at most two agents may interact at any time and may choose to change state based on this communication. Since only two agents interact at a time, the model can be viewed as a restricted form of CRNs, VASs, and Petri-Nets. Due to the limited nature of the agents, all reactions are volume-preserving with two inputs and two outputs. Reachability was proven to be PSPACE-hard in \cite{Esparza2019} via a limited version of IO Petri-Nets.

\vspace*{-.2cm}
\subsection{Our Contributions}

\begin{table}[t]
    \centering
  \begin{tabular}{ | c | c  | c  | c | c |}
  \hline
  \multicolumn{5}{|c|}{\textbf{General CRNS}} \\
  \hline
  \emph{Cons.} & \emph{Source} & \emph{Rule Size} & \emph{Membership} & \emph{Theorem}\\ \hline

  $\mathcal{O}(q)$ & $\mathcal{O}(q)$ & $(2, 2)$ & PSPACE-complete & \cite{Esparza2019} \\ \hline

  $2$ & $2$ & $(2, 2)$ &  PSPACE-complete & Thm. \ref{thm:reachPSPACE} \\  \hline
  $2$ & $2$ & $(1, 2)$ and $(2, 1)$  &  PSPACE-hard & Cor. \ref{thm:nonMonotone} \\  \hline
  $j$ & $k$ & $(1, 1)$ &  NL-hard & Thm. \ref{thm:uni} \\ \hline
  $1$ & $1$ & $(k, 1)$ &  P & Thm. \ref{thm:OneOnePoly} \\ \hline
  \hline
  \multicolumn{5}{|c|}{\textbf{Feed-Forward}} \\
  \hline
  \emph{Cons.} & \emph{Source} & \emph{Rule Size} & \emph{Membership} & \emph{Theorem}\\ \hline
  $2$ & $2$ & $(1, 2)$ or $(2, 2)$ or $(2,1)$  & NP-complete & Thm. \ref{thm:HamPath}\\ \hline
  $j$ & $1$ & No Void  & P &  Thm. \ref{thm:ff-ss-nv-P} \\ \hline
  $1$ & $k$ & No Autogenesis & P &  Thm. \ref{thm:ff-sc-na-P}  \\ \hline
  $1$ & $1$ & Any  & P & Cor. \ref{cor:ff-ss-sc-any-P}  \\ \hline
  \hline
  \multicolumn{5}{|c|}{\textbf{Feed-Forward with Void/Autogenesis Rules}} \\
  \hline
  \emph{Cons.} & \emph{Source} & \emph{Rule Size} &  \emph{Membership} & \emph{Theorem}\\ \hline
  $3$ & $0$ & $(3, 0)$  & NP-complete & Thm. \ref{thm:void3}\\ \hline
  $j$ & $0$ & $(2, 0)$  & P & Thm. \ref{thm:void2} \\ \hline
  
  \end{tabular}
  \caption{Table of reachability results. 
  Rule sizes $(a,b)$ indicates $a$ elements interact as input to produce $b$ elements as output. $q$ denotes the number of states in the Turing machine. For the membership column, only hardness was shown for two results, and membership is discussed more in those sections. } \label{tab:results}
\end{table}

In this paper, we improve on nearly all known results related to monotone volume-restricted CRNs, where each reaction preserves the volume of the system, based on several aspects of the rules. An overview of the major results is listed in Table \ref{tab:results}. The results in the three sub-tables roughly correspond to the major results in Sections \ref{sec:general}, \ref{sec:feedforward}, and \ref{sec:void}, respectively.

\para{Proper/General CRNs.}
The general reachability problem in CRNs is Ackermann-complete~\cite{czerwinski:2022:FOCS,leroux:2022:FOCS}, and thus we focus on the restriction of \emph{proper} CRNs, in which each reaction/rule preserves or decreases system volume, putting the problem in PSPACE. 
Practical considerations motivate the study of small reactions, such as bimolecular reactions in which up to 2 molecules are used as products (such as in Population Protocols). Although this was recently shown in \cite{Esparza2019}, the number of rules producing and consuming any given species was based on the size of the input.
We show that reachability is PSPACE-complete with bimolecular reactions even when at most two rules consume or produce a species (Theorem~\ref{thm:reachPSPACE}).  Thus, reachability in population protocols is PSPACE-complete under this restriction (2-source, 2-consuming).  Parameterizing systems based on their source/consuming number is a new approach introduced in this paper, and we believe our results suggest it is an important parameter for determining model complexity, especially around the boundary between values 2 and 1.

In Corollaries \ref{cor:univReach} and \ref{thm:nonMonotone}, this reduction extends to the universal reachability problem, and reachability for non-monotone volume. For production, our PSPACE-complete result is tight with regards to rule size as we show NL-completeness for $(1,1)$ rules in Theorem \ref{thm:uni}. We also provide several related smaller results.

\para{Feed-Forward CRNs.}
Feed-forward CRNs essentially permit an acyclic ordering of reactions and have important applications due as they allow for functional composition~\cite{vasic2022programming}.  We fully characterize, based on rule size, void or autogenesis rules, and the number of source/consuming rules, all CRN systems that are feed-forward, which is where reactions of the CRN permit an ordering such that products of later reactions do not occur as reactants in earlier rules \cite{chen2014rate}. We show that, without void/autogenesis rules, the feed-forward property alone moves the reachability problem into the class NP (Theorem~\ref{lem:ff-na-NP}). We show that reachability is polynomial-time solvable for a feed-forward system if it is \emph{either} 1-consuming \emph{or} 1-source and uses no \emph{void} rules (rules that produce no species) or \emph{autogenesis} rules (rules that consume no species). 
We prove that relaxing these restrictions makes reachability hard by showing that the problem in feed-forward systems is NP-complete in 2-consuming, 2-source systems (Theorem~\ref{thm:HamPath}), and if they have void or autogenesis rules.

\para{Consuming and Source.}
A $k$-consuming CRN is one where the number of rules that consume or ``use up'' a species is bound by $k$. $k$-source systems limit the number of rules that a species may be sourced from, or created by, the CRN. Non-Competitive CRNs \cite{vasic2022programming} are a special case of $1$-consuming where any rule that is consumed is not allowed to be a catalyst. Consuming and source metrics are not only interesting from a theoretical perspective since reachability is hard even for small $k$, but also from a implementation aspect. When implementing CRNs with DNA strand displacement, one method implements reactions as larger DNA complexes that bind to smaller DNA strands complexes \cite{soloveichik2010dna}. Limiting source and consuming rules make the DSD systems less complex as we are limiting the number of complexes the strands must interact with. 

\para{Void and Autogenesis Rules.}
Our final consideration is the case of systems that utilize a particularly powerful class of rules: \emph{void} rules that do not produce any species types, and \emph{autogenesis} rules which do not consume any species types. The study of such \emph{deletion-only} (or \emph{creation-only}) systems is a natural restriction of general CRNs that can both create and remove species, and the limitation to only removal could plausibly support simpler molecular implementations.  Moreover, recent work has explored using such deletion-only systems to implement general Threshold Circuits~\cite{anderson2024computing,Anderson:2024:UCNC}.  We show that reachability is NP-complete even if system reactions are all void rules or all autogenesis rules of size $(3,0)$ or $(0,3)$, respectively (Theorem~\ref{thm:void3}). We then explore the case of $(2,0)$ void rule systems and show that reachability is polynomial-time solvable (Theorem~\ref{thm:void2}). We also show a couple of special cases that are straightforward: if the volume of the input configuration is polynomial bounded (Theorem~\ref{thm:void2uni}), \emph{or} if the CRN is \emph{bipartite} (Theorem~\ref{thm:void2bi}). 

\section{Preliminaries} \label{sec:prelim}

\para{Basics.}
Let $\Lambda= s_1, s_2, \ldots, s_{|\Lambda|}$ denote some ordered alphabet of \emph{species}. A configuration over $\Lambda$ is a length-$|\Lambda|$ vector of non-negative integers, denoting the number of copies of each present species.  A \emph{rule} or \emph{reaction} has two multisets, the first containing one or more \emph{reactant} (species) used to create resulting \emph{product} (species), the second multiset.  We represent each rule as an ordered pair of configuration vectors $R=(R_r, R_p)$. $R_r$ contains the minimum counts of each reactant species necessary for reaction $R$ to occur, where reactant species are either \emph{consumed} by the rule in some count or leveraged as \emph{catalysts} (not consumed); in some cases a combination of the two.  The product vector $R_p$ has the count of each species \emph{produced} by the \emph{application} of rule $R$, effectively replacing vector $R_r$.  The species corresponding to the non-zero elements of $R_r$ and $R_p$ are termed \emph{reactants} and \emph{products} of $R$, respectively.

The \emph{application} vector of $R$ is $R_a = R_p - R_r$, which shows the net change in species counts after applying rule $R$ once.  For a configuration $C$ and rule $R$, we say $R$ is applicable to $C$ if $C[i] \geq R_r[i]$ for all $1\leq i\leq |\Lambda|$, and we define the \emph{application} of $R$ to $C$ as the configuration $C' = C + R_a$. For a set of rules $\Gamma$, a configuration $C$, and rule $R\in \Gamma$ applicable to $C$ that produces $C' = C + R_a$, we say $C \rightarrow^1_\Gamma C'$, a relation denoting that $C$ can transition to $C'$ by way of a single rule application from $\Gamma$.
We further use notation $C\rightsquigarrow_\Gamma C'$ to signify the transitive closure of $\rightarrow^1_\Gamma$ and say $C'$ is \emph{reachable} from $C$ under $\Gamma$, i.e., $C'$ can be reached by applying a sequence of applicable rules from $\Gamma$ to initial configuration $C$.  We use the following notation to depict a rule $R=(R_r, R_p)$:


$ \sum_{i=1}^{|\Lambda|} R_r[i]s_i \rightarrow \sum_{i=1}^{|\Lambda|} R_p[i]s_i$

For example, a rule turning two copies of species $H$ and one copy of species $O$ into one copy of species $W$ would be written as $2H + O \rightarrow W$.

\begin{definition}[Discrete Chemical Reaction Networks]  A discrete chemical reaction network (CRN) is an ordered pair $(\Lambda, \Gamma)$ where $\Lambda$ is an ordered alphabet of species, and $\Gamma$ is a set of rules over $\Lambda$.
\end{definition}

The primary computational problem we consider in this paper is the \emph{reachability} problem. 
We consider additional problems in the paper, such as determining if it is possible to produce a given amount of a particular species from an initial configuration of a CRN, as well as universal reachability, which asks if the target configuration is reachable for all reaction application sequences.




\begin{definition}[Reachability Problem.]  Given a CRN $(\Lambda,\Gamma$), an initial configuration $I$, and a destination configuration $D$, the \emph{Reachability Problem} is to compute whether or not $D$ is reachable from $I$ with respect to $\Gamma$.
\end{definition}

\begin{definition}[Production Problem]
Given a CRN $(\Lambda, \Gamma)$, an initial configuration $I$, a species $s_i \in \Lambda$, and a positive integer $k$, decide if there exists a reachable configuration $B$ such that $B[i] \geq k$.
\end{definition}

\begin{definition}[Universal Reachability Problem.]  Given a CRN $(\Lambda,\Gamma$), an initial configuration $I$, and a destination configuration $D$, the \emph{Universal Reachability Problem} is to compute whether  $D$ is reachable from all configurations $M$ that are reachable from $I$ with respect to $\Gamma$.
\end{definition}

\begin{figure}[t]
    \centering
    \includegraphics[width=1.\textwidth]{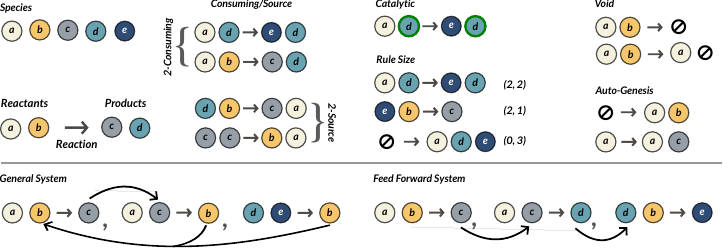}
    \caption{Example CRN rules to demonstrate the primary restrictions.}
    \label{fig:exCRN}
\end{figure}

\para{Primary Restrictions.}
We consider the reachability problem under several different restrictions defined below. Fig. \ref{fig:exCRN} provides examples of the various restrictions on the model.

\begin{definition}[Feed-Forward]  A CRN $(\Lambda, \Gamma)$ is \emph{feed-forward} if $\Gamma$ permits an ordering on the rules such that the products of any given rule never occur as reactants for earlier rules of the ordering \cite{chen2014rate}.
\end{definition}


Each rule in a system produces some species and consumes others.  The following metric places a maximum bound on the number of rules that either produce a given species ($j$-source) or consume a given species ($j$-consuming). With a 1-source, 1-consuming system, the concept is similar to free-choice Petri-Nets \cite{Zaitsev:2014:TSMCS}, but may differ greatly with only one restriction.

\begin{definition}[$j$-source, $j$-consuming]
A species $s_i$ is \emph{consumed} in rule $R=(R_r, R_p)$ if $R_r[i] > R_p[i]$, \emph{produced} if $R_r[i] < R_p[i]$, and is a \emph{catalyst} in rule $R$ if $R_r[i] = R_p[i] > 0$.
A CRN $(\Lambda , \Gamma)$ is $j$-source if for all species $s \in \Lambda$,  $s$ is produced in at most $j$ distinct rules in $\Gamma$.  A CRN $(\Lambda , \Gamma)$ is $j$-consuming if for all species $s \in \Lambda$, $s$ is consumed in at most $j$ distinct rules in $\Gamma$.  We use the terms \emph{single-source} and \emph{single-consuming} for the special cases of $1$-source and $1$-consuming CRNs, respectively.
\end{definition}

The next concept is a special class of rules that either produce nothing (void rules) or consume nothing (autogenesis rules). This could potentially be motivated by evaporation, the ability to pass through a membrane or the spontaneous appearance of ions in a vacuum.

\begin{definition}[Void and Autogenesis rules]
A rule $R=(R_r, R_p)$ is a \emph{void} rule if $R_a = R_p - R_r$ has no positive entries.  A rule is an \emph{autogenesis} rule if $R_a$ has no negative values.
\end{definition}

\para{Additional Restrictions.}
We also consider the complexity of reachability and production with respect to the \emph{size} of rules.

\begin{definition}
The \emph{size/volume} of a configuration vector $C$ is $\verb"volume"(C) = \sum C[i]$.
\end{definition}

\begin{definition}[size-$(i,j)$ rules]
A rule $R=(R_r, R_p)$ is said to be a size-$(i,j)$ rule if $(i,j) = (\verb"volume"(R_r),$ $\verb"volume"(R_p))$. A reaction is bimolecular if $i = 2$ and unimolecular if $i = 1$.
\end{definition}

\begin{definition}[Volume Decreasing, Increasing, Preserving]
A rule $R=(R_r, R_p)$ of size-$(i,j)$ is said to be volume decreasing if $i>j$, volume increasing if $i<j$, and volume preserving if $i=j$.  A CRN $(\Lambda, \Gamma)$ is said to be volume decreasing (respectively increasing, preserving) if all rules in $\Gamma$ are volume decreasing (respectively increasing, preserving). Note: In previous work, volume preserving has been called \emph{Proper}.
\end{definition}

A special subset of CRN systems studied in the literature is Population Protocols, in which agents bump into each other and adjust their state according to rules. This model is equivalent to a CRN that is limited to exactly volume 2 for both the reactants (the two agents that bump into each other) and the products (representing the two new states of the agents after the collision).

\begin{definition}[Population Protocols]
A CRN $(\Lambda, \Gamma)$ in which all rules in $\Gamma$ are size-$(2,2)$ is called a population protocol.
\end{definition}

\section{Reachability in General CRNs}\label{sec:general}

The main result of this section is PSPACE-completeness of the reachability problem with 2-source, 2-consuming, size-$(2,2)$ reactions in Theorem \ref{thm:reachPSPACE}. En route, we prove that production is PSPACE-complete with Theorem \ref{thm:prodPSPACE}. We extend this reduction in two ways.  First, in Corollary \ref{cor:univReach}, we prove our reduction holds for the universal reachability problem, and second, we show the reduction holds for non-monotonic volume with rule sizes of $(1,2)$ and $(2,1)$ in Theorem \ref{thm:nonMonotone}. We follow with several interesting, yet minor results. 

\subsection{Gadget Reconfiguration Framework}
Our main result is based on the motion planning problem through Toggle-Lock and Rotate gadgets \cite{demaine2022pspace}. Motion planning is PSPACE-hard with only a rotate gadget and any reversible gadget with interacting tunnels, a class that includes the Toggle-Lock \cite{demaine2018computational}. The motion planning problem considers two input configurations of gadgets, a start location for the agent, and a target location, and asks if the agent can reach the target location.
\footnote{In \cite{demaine2022pspace} this problem is called the reachability problem. To avoid confusion, we refer to this as the motion planning problem.}

\begin{figure}[t]
         \centering
     \begin{subfigure}[b]{0.1\textwidth}
         \centering
        \includegraphics[width=1.\textwidth]{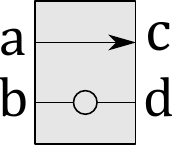}
        \caption{}
        \label{fig:unRight}
        \includegraphics[width=1.\textwidth]{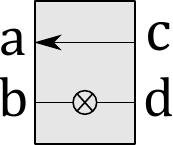}
        \caption{}
        \label{fig:lockLeft}
    \end{subfigure}
    \hspace{.0cm}
	\begin{subfigure}[b]{0.27\textwidth}
	    \centering
        $$\overrightarrow{a} + G \rightarrow \overleftarrow{c} + G'$$
        $$\overrightarrow{b} + G \rightarrow \overleftarrow{d} + G$$        
        $$\overrightarrow{c} + G \rightarrow \overleftarrow{c} + G$$
       $$\overrightarrow{d} + G \rightarrow \overleftarrow{b} + G$$
        \caption{}
        \label{fig:unlockRules}
    \end{subfigure}
    \begin{subfigure}[b]{0.27\textwidth}
	    \centering
        $$\overrightarrow{a} + G' \rightarrow \overleftarrow{a} + G'$$
        $$\overrightarrow{b} + G' \rightarrow \overleftarrow{b} + G'$$
        $$\overrightarrow{c} + G' \rightarrow \overleftarrow{a} + G$$
       $$\overrightarrow{d} + G' \rightarrow \overleftarrow{d} + G'$$
        \caption{}
        \label{fig:lockRules}
    \end{subfigure}
    \hspace{.1cm}
    \begin{subfigure}[b]{0.11\textwidth}
         \centering
        \includegraphics[width=0.8\textwidth]{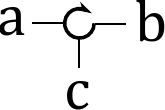}
        \caption{}
        \label{fig:rotGadget}
    \end{subfigure}
	\begin{subfigure}[b]{0.18\textwidth}
	    \centering
        $$\overrightarrow{a} + r_{\circlearrowright} \rightarrow \overleftarrow{b} + r_{\circlearrowright}$$
        $$\overrightarrow{b} + r_{\circlearrowright} \rightarrow \overleftarrow{c} + r_{\circlearrowright}$$
        $$\overrightarrow{c} + r_{\circlearrowright} \rightarrow \overleftarrow{a} + r_{\circlearrowright}$$
        \caption{}
        \label{fig:rotRules}
    \end{subfigure}
    \caption{(a) Unlocked state of a Toggle-Lock gadget represented by species $G$.
    (b) Locked state of a Toggle-Lock gadget represented by species $G'$.
    (c-d) Reactions which implement a single gadget. (c) represents a successful traversal and (d) represents the `bound-back' reactions. The arrow is incoming or outgoing from port.
    (e) The rotate gadget.
    (f) Rules for the rotate gadget.}
    \label{fig:gadget_toggle}
\end{figure}

A gadget consists of a set of labeled ports and states describing the gadget's legal traversals, which may change the state of the gadget.
The Toggle-Lock Gadget has an unlocked and locked state, shown in Figure \ref{fig:unRight} and \ref{fig:lockLeft}, respectively. The top path is directed based on the state of the gadget, and the agent must follow the direction. Traversing this path changes the state of the gadget. Traversal of the bottom tunnel can only occur in the unlocked state; this can be done in either direction and does not change the state of the gadget. The rotate gadget only has one state that sends the agent to the next port, going clockwise.
A motion planning system consists of a set of gadgets, a set of wires denoting port connections, and an initial signal location.

We are simulating a $0$-player gadget framework where the agent makes no choices. The agent is directed down the wire and turns around if the gadget is not traversable in the current state, corresponding to a deterministic model of computation. An early version of this result appeared in the short abstract \cite{popReach}, which reduced from a different gadget and used the $1$-player framework studied in \cite{ani2022traversability, demaine2018computational, demaine2020toward}. Petri-Nets reachability was shown to be equivalent to a gadget framework with unbounded agents \cite{ani2023complexity}. However, the reductions were not constant source or constant consuming. 

\para{Directed Wire.}
Two ports are connected by a wire. We label each wire with a symbol such as $a$ and include two species $\overrightarrow{a}, \overleftarrow{a}$ to denote the direction moved along the wire. We refer to these as the agent species.

\para{Toggle-Lock and Rotate Gadgets.}
We have two species for each Toggle-Lock that we call the \emph{gate catalysts}. We represent the unlocked gate using the species $G$, and the locked state with species $G'$. Traversing the toggle tunnel requires the gadget to be in the correct state for the toggle. The reaction changes the state of the gate catalyst and the agent species. The bottom tunnel can only be traversed if the gate is in the unlocked state. However, this does not change the state of the gate, so the $G$ species acts a catalyst that must be present to traverse the gadget. Figures \ref{fig:unlockRules} and \ref{fig:lockRules} shows the rules used to implement a toggle lock.

We implement rotate gadgets with a single rotate clockwise species $r_\circlearrowright$. The rotate gadget diagram is shown in Figure \ref{fig:rotGadget}. The signal state changes to the outgoing direction of the next wire in the clockwise ordering of ports using the rules shown in Figure \ref{fig:rotRules}.
Note that each reaction consumes and creates exactly one agent species yielding the following observation:

\begin{observation}\label{obs:agent}
    Any reachable configuration in the reduction only contains a single agent species.
\end{observation}

\subsection{Production and Reachability}
We prove production is PSPACE-complete using the framework described above. The target species is the agent species representing the target wire.

\begin{theorem}\label{thm:prodPSPACE}
Production in 2-source, 2-consuming preserving CRNs is PSPACE-complete with only bimolecular reactions.
\end{theorem}

\begin{proof}
    Membership in PSPACE for Proper CRNs was shown in \cite{thachuk2012space}. Given an instance of the motion planning problem with toggle-locks and rotate gadgets, create a CRN ruleset and configuration as described above.
    Our starting configuration of the CRN encodes the start location and starting states of the gadgets. The target species we wish to produce is the agent species for the target location in the motion planning problem.

    The agent may only traverse a toggle-lock gadget if the gate catalyst is in the correct state. The rotate gadget sends the signal to the correct next state.
    Once the agent reaches the target wire, the agent species representing it is produced.
    If the agent reaches the target location, then reactions apply representing the path of the agent to produce the target species.

    From Observation \ref{obs:agent}, there only exists one agent species and the reactions encode only valid traversals through the gadgets.
    If the target species is produced, then the sequence of reactions to produce it represents the path of the agent though the system of gadgets.
    Each gate catalyst is only produced and consumed in the reaction that implements the toggle tunnel.
    Each agent species is directed, so it also is only produced and consumed in one rule.

    Note that if an agent ever attempts to cross a gadget in a state that does not allow that traversal, the configuration will no longer have any valid reactions. This is the equivalent to an illegal move in the motion planning problem and the agent cannot progress, so the game is over.
\end{proof}

We extend the reduction above to work for reachability by taking advantage of the fact that the toggle-lock is reversible.
Once we reach the target species, we flip the rotate catalyst to allow the agent to move counterclockwise through a rotate gadget, allowing us to undo all the changes to the gadget states. When the agent reaches the starting location again, the system is in the initial configuration except with the opposite rotate catalyst.
Reconfiguration for the 1-player version of the motion planning framework was shown to be PSPACE-complete using a similar technique where the agent changes the state of the final gadget, then undoes all of its previous movements \cite{ani2022traversability}.
This reduction extends to the universal reachability problem since there is only one reaction at each step that can be performed.

\begin{theorem}\label{thm:reachPSPACE}
Reachability in 2-source, 2-consuming, preserving CRNs is PSPACE-complete with only bimolecular reactions.
\end{theorem}
\begin{proof}
    The reduction from Theorem \ref{thm:prodPSPACE} can be extended to show the reachability problem is PSPACE-hard as well.
    We add an additional species $r_\circlearrowleft$ which is the rotate counterclockwise catalyst.
    The target configuration is the same as the initial configuration except with the counterclockwise catalyst $r_\circlearrowleft$.
    If the target wire is $a$, we add the rule $\overrightarrow{a} + r_\circlearrowright \rightarrow \overleftarrow{a} + r_\circlearrowleft$, which changes the direction of the rotate catalyst and turns the agent around on the wire.
    Since the toggle lock gadget is reversible, there exists a sequence of reactions to return it to the initial state, the agent will undo all of its moves and return to the start configuration.
    Since the gadget system has the property of being reversible, this backwards traversal is possible.
\end{proof}

\begin{corollary}\label{cor:univReach}
    Universal reachability in 2-source, 2-consuming preserving CRNs is PSPACE-complete with only bimolecular reactions.
\end{corollary}
\begin{proof}
    From Observation \ref{obs:agent}, we know there only exists one agent state in the system at a time. Since the system is single-consuming, the agent state is only consumed in a single rule.
    These two points mean there only exists a single move sequence, so if the target configuration is reachable, it is universally reachable.
\end{proof}

\vspace{-.5cm}
\subsection{Volume Related Results}
Here, we look at several restrictions related to rule size, and consequently, volume.

\para{Non-Monotone Volume.}
We extend the reduction to utilize smaller rules. We show PSPACE-hardness of production and reachability when allowing both $(1,2)$ and $(2,1)$ rules. The CRN has both volume increasing and decreasing rules, which means it is non-monotone, and thus, these problems are not known to be in PSPACE.
To prove this, we add an intermediate species for each reaction. We replace the reaction $\overrightarrow{a} + G \rightarrow \overrightarrow{c} + G'$ with the two reactions  $\overrightarrow{a} + G \rightarrow \overrightarrow{aGc}$ and $\overrightarrow{aGc} \rightarrow \overrightarrow{c} + G'$. Although this does allow for different volumes, we believe that these rules do not provide the flexibility needed for general CRN reachability, which is Ackermann-complete. However, we do not have a proof of membership for PSPACE, so it may be feasible.

\begin{corollary}\label{thm:nonMonotone}
    Production and reachability in 2-source, 2-consuming CRNs is PSPACE-hard with rules of size $(2,1)$ and $(1,2)$.
\end{corollary}
\begin{proof}
The reduction behaves the same way as in Theorems \ref{thm:prodPSPACE} and \ref{thm:reachPSPACE}, however, here, we either have a single agent species, or a single intermediate species.  
\end{proof}

\para{Unary Encoded Volume.}
When a system is volume-increasing or volume-decreasing, the reaction sequence length is polynomial in the volume of the system, so we achieve the following theorem.

\begin{theorem}\label{thm:unaryNP}
    Reachability is in NP for volume-increasing and volume-decreasing CRNs when the volume is encoded in unary.
\end{theorem}
\begin{proof}
When the volume of a system is strictly increasing or decreasing, any reaction sequence between $I$ and $D$ is bounded by of size $\leq |I-D|$. The sequence can then be given as a `yes' certificate for reachability.
\end{proof}
\para{Unimolecular Reactions.}
Unimolecular reactions are of the form $A \rightarrow B$, i.e. preserving rules of size 1. If we are limited to only this type of reaction, production is NL-complete. The NL-hardness result works for reachability as well. 

\begin{theorem}\label{prodNLC}
    Production with rules of size $(1,1)$ is NL-complete.
\end{theorem}
\begin{proof}
Non-deterministically select a species $i$ with a positive count in the initial configuration, check if the target species is reachable from $i$, if yes then the target species is producible. Checking reachability is in NL, since the reactions can be viewed as directed edges. NL-hardness comes from Theorem \ref{thm:uni}.
\end{proof}

We only show hardness for this case. The na{\"i}ve non-deterministic algorithm does not work when the input values of the vectors are encoded in binary. We would need to track the entire configuration in-between each step. We leave exact membership for future work. 

\begin{theorem}\label{thm:uni}
    Reachability and production in CRNs is NL-hard with rules of size $(1, 1)$.
\end{theorem}
\begin{proof}
Given a directed graph $G$ we create a set of species and reactions as follows. For each node $v \in G$ we create a species. For each edge $(a, b) \in E$ we create a reaction $a \rightarrow b$.

We reduce from the directed path problem: given two nodes $s,t \in G$, does there exist a path between $s$ and $t$? Let our initial configuration be a single copy of the species representing $s$ and our target configuration be a single copy of $t$. At each step the species will represent the current node in the path to reach $t$ if and only if there exists a path.
\end{proof}


\subsection{Results Related to Single-Consuming or Single-Source}
Although the majority of the results in this section relate to rule size, we also look at rule restrictions based on the number of source and consuming rules. Our main result shows that reachability in a 2-consuming, 2-source general CRN system is PSPACE-complete. In Sections \ref{sec:feedforward} and \ref{sec:void}, we fully characterize these systems if the rules are feed-forward. Here, we address a few interesting results for CRNs that are not feed-forward.


\para{Production is NP-Hard.}
Here, we state that production is NP-hard in single-consuming CRNs. We reduce from 3-SAT by creating species for the variables and clauses, as well as creating rules to assign variables and satisfy clauses. 

\begin{theorem}
    Production in single-consuming CRNs is NP-hard.
\end{theorem}
\begin{proof}
Consider an instance of 3SAT where $X$ is the set of variables and $C$ is the set of clauses. For each $x_i \in X$ we include $x_i, \overline{x_i}, x_i^T, x_i^F$ in our species set.
The starting configuration includes a copy of each of the species $x_i, \overline{x_i}$.
For each $c_j \in C$ with $c_j = (x_a, x_b, x_c)$, we include $c_j^a, c_j^b, c_j^c, c_j^{SAT}, SAT_j$.
The starting configuration includes a copy of each species $c_j^a, c_j^b, c_j^c$ and $SAT_0$.
The target species we wish to produce is $SAT_{|C|}$.

Two assignments species $T, F$ are included. One copy of each of these species is included in the starting configuration that acts as a catalyst.
There are two rules for each $x_i$:
$T + x_i + \overline{x_i} \rightarrow T  + x_i^T + \overline{x_i}$ for assigning true,
and $F + x_i + \overline{x_i} \rightarrow F + x_i + x_i^F$ for assigning false. The $T$ and $F$ catalysts are used in order to not have two rules with the same reactants. The rules change one of the species to the true or the false species, and the other can not be created. 
We add a rule $c_j^a + x_a^T \rightarrow c_j^{SAT} + x_a^T$  for positive literals and $c_j^a + x_a^F \rightarrow c_j^{SAT} + x_a^F$ for negated literals.
To verify each clause we include the rules, $SAT_j + c_j^{SAT} \rightarrow SAT_{j + 1} $

The target $SAT_{|C|}$ is only producible by starting with $SAT_0$ and applying the rule $SAT_j + c_j^{SAT} \rightarrow SAT_{j + 1} $ exactly $|C|$ times to verify the satisfaction of each clause.
\end{proof}


\subsection{A Polynomial-Time Algorithm for Single-Source and Single-Consuming Simple CRNs}\label{sec:polyoneone}

\newcommand{\dist}{{\rm dist}}

In this section, we show the existence of a polynomial-time algorithm for single-source single-consuming simple CRNs where each rule is of the form $a_1+\cdots +a_k\rightarrow b$, and  satisfies the requirements that $k\ge 2$ and $a_1,\cdots, a_k,b$ are different species.

\para{Preliminaries.}
A CRN $(\Lambda,\Gamma$) is a $(1,1,1)$-CRN if it is  $1$-source and $1$-consuming, 
and each rule $a_1+\cdots +a_k\rightarrow b$ satisfies that $k\ge 2$ and the species $a_1,\cdots, a_k,b$ are different from each other.
A $(1,1,1)$-CRN is also called {\it single source and single consuming simple} CRN.

Let $G=(V,E)$ be a directed graph with $|V|=n$. A directed graph $G=(V,E)$ is weakly connected if the undirected graph $G'=(V, E')$ is connected, where $E'$ is the set of undirected edges $(u,v)$ with $u\rightarrow v$ as a directed edge in $G=(V, E)$. A vertex $v$ is a {\it leaf } in $G=(V,E)$ if there is one edge $u\rightarrow v$ coming to it. A vertex $v$ is an {\it almost leaf} if every edge $u\rightarrow v$ in $E$ has $u$ as a leaf. 

Let $(\Lambda,\Gamma)$ be a $(1,1,1)$-CRN. Let each species of the CRN be a vertex in a directed graph $G=(V, E)$. There is an edge $u\rightarrow v$ if there is a rule that has $u$ on the left and $v$ on the right side. We say $G=(V,E)$ is {\it derived} from the CRN $(\Lambda,\Gamma)$.

The reachability problem for a directed graph $G=(V,E)$ derived from a CRN $(\Lambda,\Gamma)$,  is to find a way to apply rules to reach configuration $(j_1,\cdots, j_n)$ from the starting configuration $(i_1,\cdots, i_n)$, where vertex $t$ has $i_t$ copies in the beginning and $j_t$ copies in the end. 

Let $G=(V,E)$ be directed a graph. We say $G$ is an \emph{almost-cycle} if it contains a unique directed cycle $C$ and every vertex has an edge directed to a vertex in the cycle $C$. 

\begin{lemma}
Let $G=(V,E)$ be derived from a $(1,1,1)$-CRN $(\Lambda,\Gamma)$ and be weakly connected. Then $G=(V,E)$ has at most one directed cycle. Furthermore, if $G=(V,E)$ does not have a directed cycle, then it is a directed tree with a root that every vertex has a directed path to. If $G=(V,E)$ does have a directed cycle, then every vertex has a directed path to a vertex in the cycle.
\end{lemma}

\begin{proof}
This is because $(\Lambda,\Gamma)$ is a $(1,1,1)$-CRN. Assume there are two different directed cycles $C_1$ and $C_2$ in $G$. We discuss two cases.
\begin{itemize}
    
    \item Case 1. There is a common vertex $v$ in both $C_1$ and $C_2$. We have that $v$ has two outgoing edges from $v$. This contradicts that $G$ is derived from a $(1,1,1)$-CRN since this implies the CRN is 2-consuming. 

    \item Case 2. There is a common vertex between $C_1$ and $C_2$. We can find a weak path connecting $C_1$ and $C_2$, where a weak path is a set of directed edges in $G$ and form a regular path when they are converted into undirected edges. One vertex on the weak path must have two outgoing edges. This contradicts that $G$ is derived from a $(1,1,1)$-CRN since this implies the CRN is 2-consuming.
\end{itemize}
Therefore, $G$ has at most one directed cycle, or a single vertex graph. If it is an almost-cycle, we are done with the proof.

Now assume that $G$ is neither an almost-cycle nor a single vertex graph. Let $v\rightarrow v_1\rightarrow\cdots \rightarrow v_k$ be a longest directed path without any vertex in the directed cycle. We have that $v_1$ is an almost leaf. Otherwise, the directed path would not be the longest. Assume that $v_1$ is generated by $v, u_1,\cdots, u_t$ in $(\Lambda,\Gamma)$ (it has a rule $v+u_1+\cdots+u_t\rightarrow v_1$). We transform $G=(V,E)$ into $G'=(V\setminus\{v, u_1,\cdots, u_t\}, E\setminus\{v\rightarrow v, u_1\rightarrow v,\cdots, u_t\rightarrow v\})$. $G'$ is derived from another $(1,1,1)$-CRN $(\Lambda',\Gamma')$ that is transformed from $(\Lambda,\Gamma)$ by removing the rule $v+u_1+\cdots+u_t\rightarrow v$ and $v,u_1,\cdots, u_t$ from $(\Lambda,\Gamma)$. This is  proven via a simple induction.
\end{proof}

\begin{lemma}\label{shrink-lemma}
Let $G=(V,E)$ be derived from a $(1,1,1)$-CRN $(\Lambda,\Gamma)$. Let $u$ be an almost leaf (not a vertex on a directed cycle) in $G=(V,E)$. Then the reachability problem for $G=(V,E)$ can be transformed into $G'=(V\setminus U, E\setminus E(U))$, where $U$ is the set of vertices $v$ entering $u$ ($(v\rightarrow u)\in E$) and $E(U)$ is the set of edges ($(v\rightarrow u)\in E$) entering $u$ in $G=(V,E)$.    
\end{lemma}

\begin{proof}
For an almost leaf $u$ not in a cycle, there is a unique way to apply the rules in order to reach the target configuration. If $u$ has $i_1$ copies in the input configuration,  $j_1$ copies in the target configuration, we must have $i_1\le j_1$. Otherwise, the target configuration is not reachable. If rule $v_1+\cdots+v_t\rightarrow u$ is the only one to generate $u$, then we have to apply these rules $j_1-i_1$ times. After applying this rule $j_1-i_1$ times, we check if the number of copies of those $v_1,\cdots, v_t$ are equal to their target values, respectively.
\end{proof}

\begin{lemma}\label{cycle-alg-lemma}
There is a polynomial-time algorithm to solve the reachability problem for an almost-cycle graph derived from a $(1,1,1)$-CRN.
\end{lemma}

\begin{proof}Let $G=(V,E)$ be an almost-cycle graph with a unique $C$ in it.
Let $v$ be a vertex in $C$. 
The vertex $v$ appears in a rule $R: v+v_1+\cdots+v_t\rightarrow u$. It is easy to see that none of $v_1,\cdots, v_t$ is the cycle $C$. If $v_i$ is in $C$, we have $u$ with both $v$ and $v_i$ going toward it in $C$, which is  a contradiction. Thus, the number of copies of each $v_i$ is decreasing. Therefore, we know the number of times the rule $R$ should be used.   For each $v$ in $C$, we compute the number of copies ($C_v$) that $v$ is consumed and the number ($P_v$) of copies $v$ are produced. The difference $C_v-P_v$ is how many times (noted $g_v$) we enter and leave $v$. 

Assume that the target configuration is reachable from the current configuration. Let it start from a vertex $v$ in the cycle $C$. If it does not finish the first $g$ iterations, we can adjust the operations until it finishes the first $g$ iterations. Assume that the broken point is at vertex $v'$.

We can decide the number of iterations to apply rules along the cycle $C$. It will be  $g=\min\{g_v\}$ if $g_v>0$ for each vertex $v$ in $C$. Without loss of generality, let $g=g_{v_1}$. 

Assume that the target configuration is reachable from the current configuration. Let it start from a vertex $v_i$ in the cycle $C$. If it does not finish the first $g$ iterations, we can adjust the operations so that it finishes the first $g$ iterations along the cycle $C$. Assume that the broken point is at vertex $v_j$. The next operation is at a vertex $v_k$ with $|j-k|>1$. We apply the rules $R_1,R_2\cdots, R_{s-1}, R_s$, where $R_s$ is the rule to generate $v_{j+1}$. The operations can be transformed to $R_s,R_1,R_2\cdots, R_{s-1}$.

After applying the rules along the cycle $C$ with $g$ iterations, we compute  $g_v'=g_v-g_{v_1}$, $C_v'=C_v-g_{v_1}$ and $P_v'=P_v-g_{v_1}$ for all vertices $v$ in $C$. We have $g_{v_1}'=0$. Let $v_1,v_2,\cdots, v_n$ be the $n$ vertices in the cycle $C$ according the direction of the cycle. For $i=2,3,\cdots, n, 1$, we apply the rule $P_{v_i}'$ times to generate $v_i$, and another rule $P_{v_i}$ times to produce $v_i$. At the end, we check if we fail or reach the target configuration.
Thus, we have a polynomial-time algorithm for the reachability of $(1,1,1)$-CRNs.
\end{proof}





\begin{theorem}\label{thm:OneOnePoly}
    There is a polynomial-time algorithm to solve the reachability problem for single-source and single-consuming simple CRNs.
\end{theorem} 

\begin{proof}
Let $(i_1,\cdots, i_n)$ be the configuration of the input species, where $i_t$ is the number of copies of the species $t$. Let $(j_1,\cdots, j_n)$ be the target configuration of species in the end of process.
 We need to find a polynomial-time solution for each weakly connected component. Without loss of generality, we assume $G$ only has one weakly connected

\begin{itemize}
    
    \item Case 1. There is no directed cycle in $G$. By Lemma~\ref{shrink-lemma}, we can shrink the graph $G$ until it becomes a trivial case.

    \item Case 2. There is one directed cycle $C$ in $G$. By Lemma~\ref{shrink-lemma}, we can shrink the graph $G$ until it becomes an almost-cycle, and then by Lemma~\ref{cycle-alg-lemma}, we know reachability is decidable in polynomial-time.

\end{itemize}

Therefore, we have a polynomial-time algorithm for the reachability problem for $(1,1,1)$-CRNs.
\end{proof}






\section{Reachability in Feed-Forward CRNs}\label{sec:feedforward}
Having established PSPACE-completeness for general CRNs, we consider the feed-forward restriction, in which rule sets do not have cycles. Feed-forward CRNs are motivated in that they allow functional composition of CRNs~\cite{vasic2022programming}. We characterize the complexity of reachability for feed-forward CRNs under the assumption that the system does not contain either void or autogenesis rules, leaving a focused consideration of void and autogenesis rules for Section~\ref{sec:void}. 

We first show NP-completeness of feed-forward systems in Section~\ref{sec:ffnpc}, even in the case of size-$(2,2)$ rules (i.e., bimolecular rules / Population Protocols), while at the same time being only $2$-source and $2$-consuming. We then show a polynomial-time solution to reachability in Section~\ref{sec:ffpoly} for any feed-forward system that is either $1$-source or $1$-consuming, thus giving a complete characterization of feed-forward reachability.

\subsection{NP-completeness for Bimolecular Reactions}\label{sec:ffnpc}
In this section, we show that reachability (and production) in feed-forward systems is NP-complete for bimolecular reactions (rules of size at most $(2,2)$), even when the CRN is 2-source and 2-consuming. This hardness result is tight because a decrease to either 1-source or 1-consuming, as shown in Section~\ref{sec:ffpoly}, implies a polynomial time solution to reachability. We start with proof of membership in NP, followed by NP-hardness from a reduction from the Hamiltonian Path problem.

\para{NP Membership.}
A key property of feed-forward systems is that any sequence of rule applications can be reordered such that all system rules are applied consecutively. This new ordering is still a valid sequence of applicable rules that reaches the same final configuration.

\begin{definition}[Ordered Application] A sequence of reactions is an ordered application if all the applications of any given rule take place right after each other in a contiguous sequence.
An example of this is $R_1, R_1, \ldots, R_1,$ $R_2, \ldots, R_2,$ $R_3$.
\end{definition}

\begin{lemma}\label{lem:swap}
    Let $C=(\Lambda , \Gamma)$ be a feed-forward CRN with a feed-forward ordering $F = \{R_0, R_1, \ldots, R_{|R|-1}\}$ over $\Gamma$. Given configurations $c$, $c'$, and a sequence of reactions $S = \{ \ldots, R_j, R_i, \ldots \}$ in $\Gamma$ that converts $c$ to $c'$ where $i < j$, then the sequence $S' = \{ \ldots, R_i, R_j, \ldots \}$, i.e., the sequence obtained by swapping the two rules $R_i$ and $R_j$, also transforms $c \rightarrow c'$.
\end{lemma}
\begin{proof}
Let $X$ denote the configuration obtained by applying the rules of $S$ up to just before the application of rule $R_j$. Let $R_j = (R^j_r, R^j_p)$ and $R_i = (R^i_r, R^i_p)$. As $S$ is a valid sequence of rule applications, $X - R^j_r$ is a non-negative vector. Due to the feed-forward ordering $F$, the reactants $R^i_r$ do not occur as products in $R^j_p$, and thus $X-R^j_r - R^i_r$ is also non-negative, implying that rule $R_i$ and $R_j$ can be applied in either order.
\end{proof}


\begin{corollary}\label{cor:ordered}
    A configuration $D$ is reachable from a configuration $I$ with a feed-forward CRN if and only if $D$ is reachable from $I$ by an ordered application of rules.
\end{corollary}

\begin{lemma}\label{lem:ff-na-NP}
    The reachability problem is in NP for feed-forward CRNs that do not use autogenesis rules.
\end{lemma}
\begin{proof}
We utilize the ordered application from Corollary \ref{cor:ordered} as a polynomial-sized certificate.  Denote this certificate as $A = \langle a_1, a_2, a_3, \ldots, a_k \rangle$ where each $a_i$ is an nonnegative  integer denoting the number of applications of rule $R_i$ in the feed-forward ordering. While sequence $A$ could include a large (exponential) number of total rule applications, we bound the number of applications by bounding the volume of the system at any given point during the application of these rules, which both shows that the sequence can be represented in size polynomial in the input size $n$ of the reachability problem, and verified in polynomial time, thus showing membership in NP.

Let $V_i$ denote the volume of the CRN after the complete application of all $a_i$ instances of rule $R_i$ in the feed-forward ordering, with $V_0 = V(I)$ denoting the volume of the initial given configuration $I$.  Note that since there are no autogenesis rules, we know that $a_i \leq V_{i-1}$, since each rule must exhaust at least one count of some species in the system.  Let $V(p_i)$ denote the volume of the product of the $i^{th}$ rule in the feed-forward ordering, and let $V(p)$ denote the largest $V(p_i)$ plus 1, noting that each application of $R_i$ increases the system volume by at most $V(p_i)\leq V(p)$.  Thus, $V_i \leq V_{i-1} + a_i V(p_i) \leq V_{i-1} + V_{i-1}V(p_i) \leq V_{i-1}V(p)$.  This recurrence equation solves to $V_i \leq V(0)V(p)^i$.  If we let $n$ denote the input size to the reachability problem, we know that $V(0) \leq 2^n$, and $V(p)\leq 2^n$, and therefore $V_i \leq 2^n\cdot 2^{i\cdot n} \leq 2^{n^2+n}$.  Thus, the total volume of the system at any point, as well as each $a_i$, can be represented in a polynomial number of bits in $n$. We make a nondeterministic path to guess nonnegative integers $a_1, a_2,\cdots a_n$ in the $[0, 2^{n^2+n}]$, verify the system has enough reactants to support $a_i$ applications of each rules $r_i$ in the sequence, and compute the configuration $C=C_0+a_1H_1+a_2H_2\cdots +a_kH_k$, where $C_0$ is the initial configuration vector, and $H_i$ is the application vector of rule $R_i$ (see Section~\ref{sec:prelim}). Finally, we check if $C$ is the same as the target configuration. Since all integers involved in these calculations are stored using a polynomial number of bits, this computation can be performed in polynomial time.
\end{proof}

This NP membership proof can be extended to the alternate case in which a rule set contains no void rules:

\begin{definition}\label{def:reverse}
For a rule set $\Gamma$, let $\overleftarrow{\Gamma}$ be the \emph{reverse} of $\Gamma$ where $\overleftarrow{\Gamma} = \{ (a, b) | (b, a)\in \Gamma\}$.
\end{definition}

\begin{lemma}\label{lem:reverse}
For any two configurations $A,B$ and rule set $\Gamma$, $B$ is reachable from $A$ in $\Gamma$ if and only if $A$ is reachable from $B$ in $\overleftarrow{\Gamma}$.
\end{lemma}
\begin{proof}
This follows from the straigthforward observation that for any two configurations $A'$ and $B'$, $A' \rightarrow^1_\Gamma B'$ if and only if $B' \rightarrow^1_{\tiny\overleftarrow{\Gamma}} A'$.
\end{proof}

\begin{lemma}\label{lem:ff-nvoid-NP}
    The reachability problem is in NP for feed-forward CRNs that do not use void rules.
\end{lemma}
\begin{proof}
This follows from Lemma~\ref{lem:ff-na-NP} and Lemma~\ref{lem:reverse}.    
\end{proof}

%
%
%
%

\begin{figure}[t]
    \vspace{-.2cm}
	\centering
		\begin{subfigure}{0.31\textwidth}
			\centering
			\includegraphics[width=0.85\textwidth]{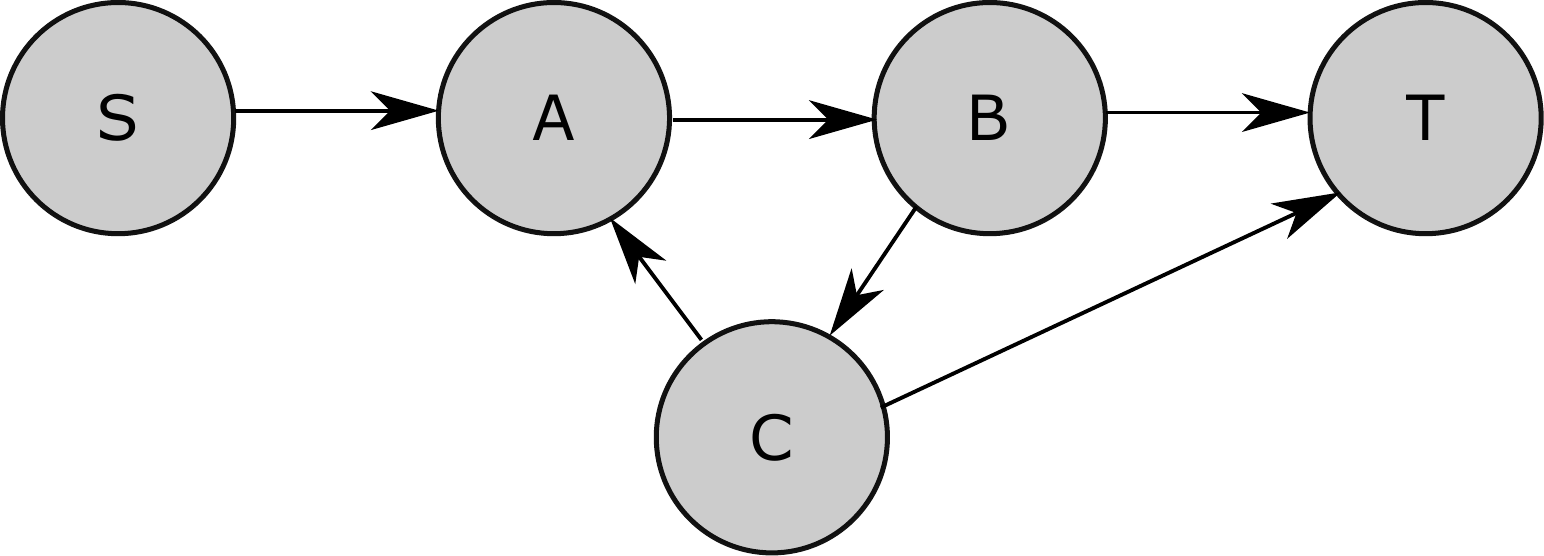}
			\caption{Graph}
            \label{fig:hampathEx}
		\end{subfigure}
		\begin{subfigure}{0.31\textwidth}
			\centering
			\includegraphics[width=0.85\textwidth]{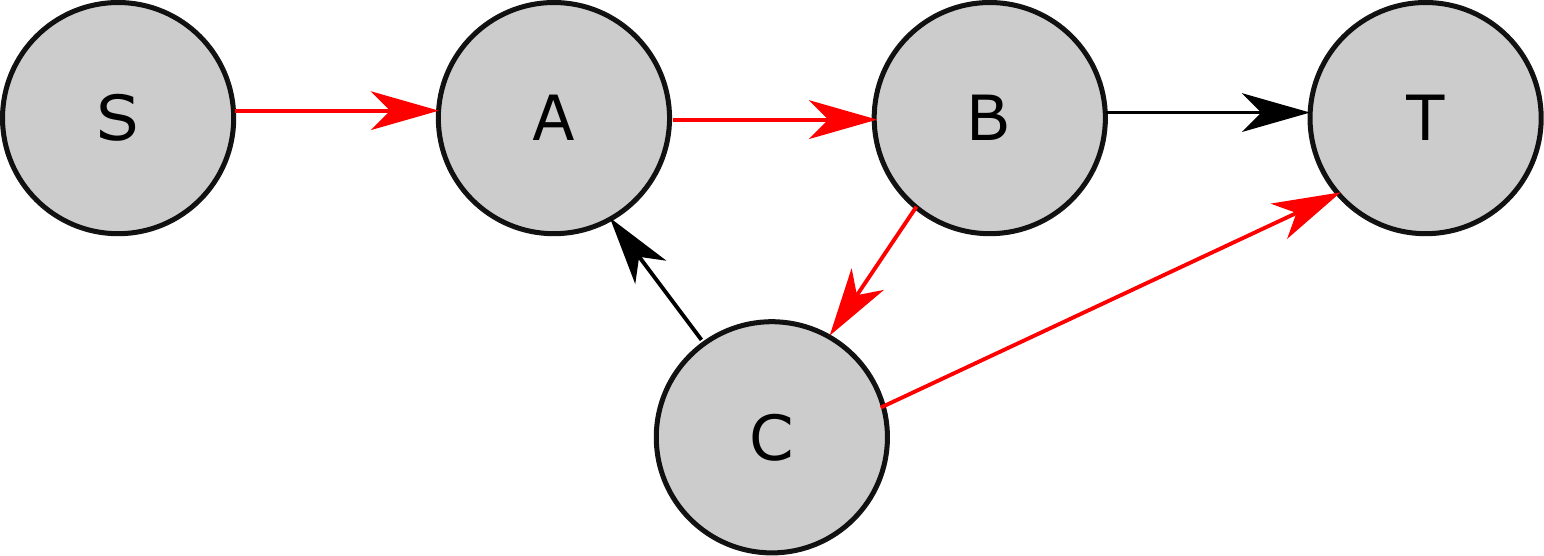}
			\caption{Hamiltonian Path}
		\end{subfigure}
	\begin{subfigure}{0.31\textwidth}
		\centering
		\begin{tabular}{c @{} c c c c c @{} c}
		$\{$ & $S^*_0$ & $A$ & $B$ & $C$ & $T$ & $\}$\\
		$\{$ & $S^v$ & $A^*_1$ & $B$ & $C$ & $T$ & $\}$\\
		$\{$ & $S^v$ & $A^v$ & $B^*_2$ & $C$ & $T$ & $\}$\\
		$\{$ & $S^v$ & $A^v$ & $B^v$ & $C^*_3$ & $T$ & $\}$\\
		$\{$ & $S^v$ & $A^v$ & $B^v$ & $C^v$ & $T^4$ & $\}$\\
		\end{tabular}
		\caption{Configurations}
	\end{subfigure}
    \vspace{-.2cm}
	\caption{Our starting configuration $c = \{ S^*_0, A, B, C, T\}$. Our goal configuration is $c' = \{ S^v, A^v, B^v, C^v, T^{4 } \}$. Each vertex must be changed to the visited state to reach the target, and the $T$ must be the last vertex. }
    \label{fig:graph}
    \vspace{-.5cm}
\end{figure}

\para{NP-Hardness.}
We now show the reachability problem is NP-complete for feed-forward CRNs even for size $(2,2)$-rules (bimolecular reactions, Population Protocols) and for 2-source, 2-consuming systems. We show this by a reduction from the Directed Hamiltonian Path problem with vertices of in-degree and out-degree of at most $2$ \cite{plesnik1979np}. We reduce from bounded degree in order to achieve bounded source/consuming. For each vertex $X$ in the graph $G=(V,E)$, we include $2 + |V|$ states: an initial state $X$, a visited state $X^v$, and $|V|$ signal states $X^*_i$. The additional signal states are added so the system is feed-forward. An example reduction is shown in Figure \ref{fig:graph}. We encode the edges of the example graph in the rules as follows,

\vspace*{.15cm}

\hspace{-.8cm}
$R = \Biggl\{$ 
\begin{tabular}{@{}c | c | c@{}}
$S^*_i + A \rightarrow S^v + A^*_{i+1}$ & $B^*_i + C \rightarrow B^v + C^*_{i+1}$ & $B^*_i + T \rightarrow B^v + T^*_{i+1}$\\
$A^*_i + B \rightarrow A^v + B^*_{i+1}$ &  $C^*_i + A \rightarrow C^v + A^*_{i+1}$ & $C^*_i + T \rightarrow C^v + T^*_{i+1}$\\
\end{tabular}$\Biggl\}$

\vspace*{.15cm}

\begin{definition}[HAMPATH]
Given a graph $G = \{V, E\}$ and two nodes $s, t \in V$, does there exist a path from $s$ to $t$ that visits each node precisely once?
\end{definition}

Given this reduction,  any Hamiltonian path of graph $G$ has a corresponding sequence of rules that end with every vertex, other than $T$, represented with the \emph{visited} state, and $T$ represented with the signal state matching the count of the vertices. Conversely, the only way to reach such a configuration corresponds directly to a Hamiltonian path of $G$ from $S$ to $T$, yielding Theorem \ref{thm:HamPath}. With minor modifications, the reduction can be adapted to achieve the corollaries below.


\begin{theorem}\label{thm:HamPath}
Reachability is NP-complete for feed-forward CRNs with size $(2,2)$ rules that are 2-source and 2-consuming.
\end{theorem}
\begin{proof}
If there exists a Hamiltonian path $P$ in $G$, there exists a sequence of rules $r_P$, each moving the $^*$ to a subsequent agent matching the sequence of vertices in $P$ and further setting the previous agent to the visited state. Such a rule sequence ends with a single visited state for each vertex in the graph other than $T$ and a $T_|V|$ state.

Conversely, if a sequence of rules $r$ that results in the target configuration exists, then a Hamilton path exists as each rule changes the agent location to the visited state, which can no longer change state, and moves the $^*$ to the next agent. The sequence of $^*$ species in the configuration represents the Hamilton path.

Finally, note that this system is feed-forward since an agent starts in the initial state, changes to a signal state, then to the visited state. We can ensure the ordering since the signal states are numbered.
\end{proof}


\begin{corollary}
Reachability is NP-complete for feed-forward CRNs that are 2-source and 2-consuming with all rules of size $(2,1)$.
\end{corollary}
\begin{proof}
This is obtained by simply removing the visited states in the previous reduction.
\end{proof}

\begin{corollary}
Reachability is NP-complete for feed-forward CRNs that are 2-source and 2-consuming and with all rules of size $(1,2)$.
\end{corollary}
\begin{proof}
This follows from the above corollary combined with Lemma~\ref{lem:reverse} (defined in Section \ref{sec:ffpoly}).
\end{proof}

\begin{corollary}
Production is NP-complete for feed-forward CRNs with either size $(2,2)$ rules, size $(2,1)$ rules, or size $(1,2)$ rules, that are 2-source and 2-consuming.
\end{corollary}
\begin{proof}
In the previous reduction, the species $T_{|V|}$ is only producible if it is reached after reaching all other species, implying the previous reduction above holds for the production problem.
\end{proof}

\para{Species-based Restrictions.}
For a CRN to be $j$-consuming, all species must be consumed in only $j$ rules. We can relax this restriction to be $j$-consuming/$k$-source per species, meaning a rule is consumed in only $j$ rules OR produced in only $k$ rules. Each species is still bounded in one of the ways, but not both. We then show reachability, with the species-based restriction of k-consuming/1-source, is NP-hard by a reduction from the 3-Dimensional Matching (3DM) problem.

\begin{definition}[Three Dimensional Matching Problem (3DM)]\label{def:3dm}
The 3DM problem takes as input a hypergraph $H=(X, Y, Z, T)$ where $X, Y, Z$ are three disjoint sets, and $T \subseteq X \times Y \times Z$ is a set of \emph{hyperedges}. The output is whether or not there exists a subset of $T$ that covers all vertices in $H$ without any overlap.
\end{definition}

\begin{corollary}\label{thm:3dmspec}
Reachability in CRNs with each species being $k$-consuming/1-source or 1-consuming/$k$-source is NP-complete even with only one species being different than the others and the system being feed-forward without void/autogenesis rules.
\end{corollary}

\begin{proof}
    We reduce from the 3DM problem, which is hard even when each vertex is covered by at most 3 hyper edges. Let $H = (X, Y, Z, T)$ be an input to the 3DM problem. From this, create an input to the reachability problem as follows. Let $\Lambda = \{ S_v | v\in X\bigcup Y \bigcup Z \} \bigcup \{a\}$, and let $\Gamma = \{ S_x + S_y + S_z \rightarrow a | (x,y,z) \in T\}$. The initial configuration $I$ is the configuration in which each species has count $1$ except species $a$ with count 0. Let $D$ be the configuration in which each species has count $0$ except $a$, which has count $|X|=|Y|=|Z|$. Then $D$ is reachable from $I$ under $(\Lambda,\Gamma)$ if and only if $H$ has a three-dimensional matching.

    Species $a$ is never consumed, but is produced by all $k=|X|=|Y|=|Z|$ of the rules. Thus, it is 0-consuming/$k$-source. All other species are never produced, and are consumed in $3$ different rules. Thus, they are 3-consuming/0-source.
    Finally, since species $A$ is never consumed, any ordering of the rules is feed-forward, and there are no void or autogenesis rules used. By Lemma \ref{lem:ff-na-NP}, the problem is in NP.
\end{proof}

\subsection{Feed-Forward, Single-Consuming/Single-Source}\label{sec:ffpoly}
In this section, we establish Theorem~\ref{thm:ff-ss-nv-P} that shows the reachability problem is polynomial-time solvable for feed-forward, single-source rule sets that do not use void rules. We then extend this into Theorem~\ref{thm:ff-sc-na-P} to show that reachability in feed-forward, single-consuming systems without autogenesis rules is also polynomial time solvable. Lastly, we give Corollary~\ref{cor:ff-ss-sc-any-P} that states the reachability problem for feed-forward, single-source, and single-consuming rule sets is polynomial-time solvable. To achieve these results, we first introduce some machinery upon which our algorithms are based.

\begin{definition}[Leaf Rule.]
A rule $R\in \Gamma$ is a \emph{leaf rule} for $\Gamma$ if the products of $R$ do not occur as reactants within any \emph{other} rule of the system $\Gamma$ (however, the products of a leaf rule may occur as reactants within the same leaf rule). A leaf rule could also be a void rule.
\end{definition}

\begin{lemma}\label{lem:ffHasLeaf}
A feed-forward, non-empty rule set has at least one leaf rule.
\end{lemma}

\begin{proof}
The final rule in the feed-forward ordering must be a leaf rule, as its product may not occur as a reactant for any previous rule of the system, which includes all rules other than itself.
\end{proof}

\begin{lemma}\label{lem:ff-ss-nv-closedSubset}
    If $\Gamma$ is a feed-forward, single-source rule set without void rules, then so is any subset of $\Gamma$.
\end{lemma}

\begin{proof}
This follows from the definitions of feed-forward, single-source, and void rules.
\end{proof}

\begin{definition}[Pruned Configuration.]
Consider a configuration $I$, a target configuration $D$, and feed-forward rule set $\Gamma$ with non-void leaf rule $R=(R_r,R_p)$. If there exists a non-negative integer $x$ such that $D(i) \setminus xR_a(i) = I(i)$,
for all $i$ where $R_p(i) \neq 0$, and $R$ is applicable to $D \setminus xR_a$, we say that $D$ is prunable towards $I$ with respect to rule $R$, and define the \emph{pruning} of $D$ towards $I$ with respect to $R$ to be the configuration $\texttt{Prune}(D,I,R) = D \setminus xR_a$. If no such non-negative integer $x$ exists, then we say that $D$ is inconsistent with $I$ for rule $R$. Note that the integer $x$, and thus the configuration $\texttt{Prune}(D, I, R) = D \setminus xR_a$, are unique as long as $R$ is not a void leaf rule.
\end{definition}

\begin{lemma}\label{lem:inconsistentNo}
    If a configuration $D$ is inconsistent with a configuration $I$ for any leaf rule $R\in\Gamma$, then $D$ is not reachable from $I$ with a rule set $\Gamma$.
\end{lemma}

\begin{proof}
As $R$ is a leaf rule, the counts of the product species in rule $R$ are only affected by rule $R$ among the rules of $\Gamma$. Therefore, if $D$ is reachable from $I$, it must be possible to generate the counts of these species specified by $D$ by some number of applications $x$ of rule $R$. If no such integer exists, which is the definition of inconsistent, then the species counts for $R$'s products cannot equal the counts specified by $D$, making $D$ unreachable.
\end{proof}

\begin{lemma}\label{lem:pruneRecurse}
    For a rule set $\Gamma$ and configuration $D$ that is consistent with $I$ for non-void leaf rule $R\in \Gamma$, then $D$ is reachable from $I$ with a rule set $\Gamma$ if and only if $D'=\texttt{Prune}(D, I, R)$ is reachable from configuration $I$ with a rule set $\Gamma \setminus R$.
\end{lemma}

\begin{proof}
We first show that if $D'=\texttt{Prune}(D, I, R)$ is reachable with a rule set $\Gamma \setminus R$, then so is $D$ with a rule set $\Gamma$. This is because once $D'$ is reached, we know from the definition of $D'=\texttt{Prune}(D, I, R)$ that there exists a non-negative integer $x$ such that $x$ applications of rule $R$ to configuration $D'$ yields $D$, implying that $D$ is reachable.

For the other direction, suppose we can reach $D$. From the sequence of rule applications that reaches $D$, there must be exactly some non-negative integer $x$ applications of rule $R$. Create a modified sequence of configurations by omitting these $x$ operations, and you get configuration $D'=\texttt{Prune}(D, I, R)$ by application of rules from $\Gamma \setminus R$, meaning $D'$ is reachable from $\Gamma \setminus R$.
\end{proof}



We now consider the case of reachability in feed-forward, single-consuming systems without autogenesis rules.  Our approach is to \emph{reverse} the given rule set $\Gamma$ and apply our algorithm for Theorem~\ref{thm:ff-ss-nv-P}.

\begin{theorem}\label{thm:ff-ss-nv-P}
    The reachability problem is solvable in polynomial time for a rule set $\Gamma$ that is feed-forward, single-source, and without void rules.
\end{theorem}
\begin{proof}
Let $I$ denote the starting configuration, $D$ denote the destination configuration, and $(\Lambda,\Gamma)$ be a feed-forward CRN without void rules for a given reachability instance. The following recursive algorithm solves reachability for a feed-forward, single-source rule set with no void rules. 

As a base case, if the input system has $0$ rules, then $D$ is reachable if and only if $I=D$. Otherwise, identify a leaf rule $R$ from $\Gamma$, which must exist by Lemma~\ref{lem:ffHasLeaf}. Check if $D$ is consistent with $I$ for rule $R$. If not, return false, which is the correct answer by Lemma~\ref{lem:inconsistentNo}. If it is, then let $D' = \texttt{Prune}(D, I, R)$ denote the pruning of $D$ towards $I$ for rule $R$ and return the result of recursively solving reachability with initial configuration $I$, ruleset $\Gamma \setminus R$, and destination configuration $D'$, which is a valid input to this algorithm as $\Gamma \setminus R$ is assured to be a feed-forward, single-source rule set without void rules by Lemma~\ref{lem:ff-ss-nv-closedSubset}, and is assured to yield the correct result by Lemma~\ref{lem:pruneRecurse}. In total, this algorithm executes $|\Gamma|$ prune operations.  Further, as the system is both feed-forward and without void rules, a polynomial number of prune operations preserves the property that the count of any system species after each pruning is small enough to be represented in only a polynomial number of bits (the argument for this is the same as the proof of Lemma~\ref{lem:ff-na-NP}).  Therefore, the polynomial number of arithmetic operations used to compute a prune can each be completed in polynomial time, and so each prune can be performed in polynomial time, and therefore the algorithm overall finishes in polynomial time.
\end{proof}



\begin{theorem}\label{thm:ff-sc-na-P}
    The reachability problem is solvable in polynomial time for a ruleset $\Gamma$ that is feed-forward, single-consuming, and without autogenesis rules.
\end{theorem}
\begin{proof}
Given a CRN $(\Lambda, \Gamma)$ where $\Gamma$ is feed-forward, single-consuming, and without autogenesis rules, along with initial configuration $I$, and destination $D$, generate rule set $\overleftarrow{\Gamma}$.  Note that $\overleftarrow{\Gamma}$ must be single-source as $\Gamma$ is single-consuming, and must have no void rules since $\Gamma$ has no autogenesis rules, and must be feed-forward since $\Gamma$ is feed-forward.  We can therefore determine if $I$ is reachable from $D$ under $\overleftarrow{\Gamma}$ in polynomial time by Theorem~\ref{thm:ff-ss-nv-P}, which gives the answer to our original reachability problem by Lemma~\ref{lem:reverse}.
\end{proof}



\begin{corollary}\label{cor:ff-ss-sc-any-P}
    The reachability problem is solvable in polynomial time for a rule set $\Gamma$ that is feed-forward, 1-source, and 1-consuming with no further restrictions on the rule set.
\end{corollary}
\begin{proof} 
Let $I$ denote the initial configuration, let $D$ denote the target configuration, and let $(\Lambda, \Gamma)$ be a feed-forward, 1-source, and 1-consuming CRN for a given reachability instance. With a single-consuming rule set $\Gamma$, any rule $R$, void or otherwise, may be used to prune $D$ if there exists some non-negative integer $x$ such that $D(i) \setminus xR_a(i) = I(i)$. By the definition of single-consuming, the integer $x$ and the configuration $\texttt{Prune}(D,I,R)=D(i) \setminus xR_a(i)$ remain unique.

It follows that the recursive algorithm used in Theorem~\ref{thm:ff-ss-nv-P} solves reachability for a feed-forward, single-source, and single-consuming rule set by allowing void rules when pruning.
\end{proof}

\section{Void and Autogenesis Rules}\label{sec:void}
In our consideration of feed-forward CRNs, we omitted two classes of rules: \emph{void} rules that consume reactants without creating any products and \emph{autogenesis} rules that create products without consuming any reactants. One reason for separating these rules is that their lack of conservation of mass might mean they are not feasible in some experimental settings. Another important reason is that their inclusion alone substantially impacts the complexity of problems such as reachability. 

Here, we explore reachability in the scenario where \emph{all} rules are void rules or, conversely, all rules are autogenesis rules. We show that void rules (or autogenesis rules) alone imply the NP-completeness of reachability, even if such systems are both feed-forward and 0-source. We specifically show NP-completeness for size $(3,0)$ void rules. 

We then explore the complexity of reachability with size $(2,0)$ void rules and show that there always exists a polynomial solution based on $b$-matching \cite{b-matching_runtime}. We give another polynomial time solution when the system's volume is encoded in unary, or when the volume is encoded in binary and it is a restricted class of \emph{bipartite} $(2,0)$ CRNs. We note by Definition \ref{def:reverse}, that we may prove results for void only rules, and they are equivalent for autogenesis rules. 


\para{Size $(3,0)$ Void Rules / $(0,3)$ Autogenesis Rules.} We show that reachability is NP-complete by a reduction from 3-Dimensional Matching (3DM) (Definition \ref{def:3dm}).



\begin{theorem}\label{thm:void3}
Reachability for CRNs with only rules of size $(3, 0)$ is NP-complete.
\end{theorem}
\begin{proof}
We reduce from the 3DM problem. Let $H = (X, Y, Z, T)$ be an input to the 3DM problem. From this, create an input to the reachability problem as follows. Let $\Lambda = \{ S_v | v\in X\bigcup Y \bigcup Z \}$, and let $\Gamma = \{ S_x + S_y + S_z \rightarrow \varnothing | (x,y,z) \in T\}$. Let configuration $I$ be the configuration in which each species has count 1 and let $D$ be the configuration in which each species has count $0$. Then $D$ is reachable from $I$ under $(\Lambda,\Gamma)$ if and only if $H$ has a three-dimensional matching.
\end{proof}


\begin{corollary}
Reachability for CRNs with only rules of size $(0, 3)$ is NP-complete.
\end{corollary}

\begin{proof}
We show this by reduction from the reachability problem with size $(3, 0)$ rules. Consider an instance of the reachability problem with an input of a CRN $(\Lambda,\Gamma)$, an initial configuration $I$, and a destination configuration $D$ in which rules in $\Gamma$ are exclusively sized $(3, 0)$ void rules. Let $\Lambda' = \Lambda$, $\Gamma' = \overleftarrow{\Gamma}$, initial configuration $I' = D$, and target configuration $D' = I$. Observe that $\Gamma'$ consists of rules only of size $(0, 3)$. By Lemma \ref{lem:reverse}, $D'$ is reachable from $I'$ under ruleset $\Gamma'$ if and only if $D$ is reachable from $I$ under ruleset $\Gamma$.
\end{proof}

\para{Size $(2,0)$ rules with Unary Encoding.} 
As with $(3,0)$ rules, $(2, 0)$ may also be reduced by matching. This time bipartite. 

\begin{theorem}\label{thm:void2uni}
    Reachability in CRNs is in P with rules of size $(2, 0)$ if configuration counts are encoded in unary.
\end{theorem}
\begin{proof}
We show this by reducing reachability in this scenario to the two-dimensional matching problem, which has an established polynomial-time solution. We first consider the configuration $X=I-D$, creating a graph from this configuration. For each non-zero count species $X(i) >0$, $X(i)$ vertices are added to the graph of type $i$. For each rule $i+j \rightarrow \emptyset$, we add edges to the graph connecting all vertices of type $i$ to all vertices of type $j$. Then we have that $X$ can reach the empty configuration if and only if the created graph has a perfect two-dimensional matching, and thus $I$ reaches $D$ if and only if such a matching exists.
\end{proof}

\para{Bipartite CRNs with $(2,0)$ rules.}
We now consider $(2,0)$ with binary encoded species counts, which permits a potentially exponential configuration volume, making the algorithm of Theorem~\ref{thm:void2uni} no longer polynomial time. In this scenario, we consider a new restriction in which the CRN rules are bipartite:

\begin{definition}\label{BipartiteCRNdef}
A bipartite CRN $(\Lambda, \Gamma)$ is one in which the species $\Lambda$ can be partitioned into two disjoint sets $\Lambda_1$ and $\Lambda_2$ such that for each rule $R \in \Gamma$, there are at most 2 reactants of $R$ and they do not occur within the same partition of $\Lambda$.
\end{definition}

Determining if a CRN is bipartite can be solved in polynomial time by a bipartite graph detection algorithm. If the CRN is bipartite, we reduce the problem to the maximum flow problem.
Although this algorithm only works with bipartite CRNs, we conjecture that the problem is in $P$, and leave it as an important open question related to general matching in weighted graphs. 


\begin{theorem}\label{thm:void2bi}
Reachability is polynomial-time solvable for bipartite CRNs with $(2,0)$ rules.
\end{theorem}
\begin{proof}
We show this by reducing reachability for a $(2,0)$ rule bipartite CRN to the maximum network flow problem. Consider an input $(2,0)$-rule bipartite CRN $(\Lambda, \Gamma)$ with partitions $\Lambda_1$ and $\Lambda_2$, input configuration $I$, and output configuration $D$. From this, generate a max-flow instance as follows: for each $s\in \Lambda$, let the network contain a corresponding vertex $v_s$. For each rule, $a+b \rightarrow \emptyset \in \Gamma$, add an infinite capacity edge between vertices $v_a$ and $v_b$. For each $x_i \in \Lambda_1$, add an edge from the source vertex to vertex $v_{x_i}$ of capacity $I[i] - D[i]$, and for each $y_i \in \Lambda_2$, add an edge from $v_{y_i}$ to the sink vertex of capacity $I[i] - D[i]$. The maximum-flow of this network is equal to the configuration volume $I-D$ if and only if $D$ is reachable from $I$ under $\Gamma$, and therefore reachability can be computed in polynomial time using the Edmonds-Karp maximum-flow algorithm.
\end{proof}


\para{General $(2,0)$ rules.}
For our results here, we rely on the $b$-matching problem, which is a generalization of matching. This  takes the form of a traditional matching when all $b$-values are 1 and an uncapacitated $b$-matching occurs when all edges capacities are assigned $u(e) = \infty$. 

\begin{definition}[$b$-matching]
Given a graph $G = (V, E)$ and some edge capacity function $u:E \rightarrow \mathbb{N}\cup\{\infty\}$ and a $b$-value function $b:V \rightarrow \mathbb{N}$, find a maximum assignment $f:E \rightarrow \mathbb{N}$ s.t. $f(e) \leq u(e)$ for all $e \in E$ and $\sum_{e \in \delta(v)}f(e) \leq b(v)$ for all $v \in V$. We call it a perfect $b$-matching if $\sum_{e \in \delta(v)}f(e) = b(v)$ for all $v \in V$.
\end{definition}

The runtime of the maximum $b$-matching problem is strictly polynomial and runs in $O(|V|^2 \log (|V|)(|E| + |V| \log (|V|)))$ on a graph $G=(V,E)$ \cite{b-matching_runtime}. 
With a polynomial runtime for $b$-matching, we note  the following lemma and present the theorem.

\begin{lemma}\label{lem:config}
Given a $(2,0)$ void rule CRN $\zeta$.
For any pair of configurations $C_a$ and $C_b$, $C_b$ is reachable from $C_a$ with system $\zeta$ if and only if the empty configuration $C_0$ (the configuration with all 0 entries) is reachable from $C_a - C_b$ with system $\zeta$.
\end{lemma}
\begin{proof}
This follows from the fact that $\zeta$ only deletes species, and does not use catalysts.
\end{proof}

\begin{theorem}\label{thm:void2}
    Reachability for CRNs with binary encoded species with only rules of size (2, 0) is solvable in $O(|\Lambda|^2 \log (|\Lambda|)(|\Gamma| +| \Lambda| \log (|\Lambda|)))$ where $\Lambda$ is the set of species and $\Gamma$ is the set of rules.
\end{theorem}
\begin{proof}

We can reduce an instance of $(2,0)$ reachability into an instance of the maximum $b$-matching problem. 
Based on Lemma \ref{lem:config}, create the $C_d = C_b - C_a$ configuration. We then create a $b$-matching instance as follows. Turn every species $\lambda_i \in \Lambda$ into a vertex $v_i$. For rules $R \in \Gamma$ of form $s_i + s_j \rightarrow \emptyset$, create an undirected edge $(v_i,v_j)$ to represent it in our graph. Set the $b$-values for each of the vertices $b(v_i) = C_d[i]$ and assign the edge capacity function $u(e) = \infty$ for all $e \in E$. We then run the maximum $b$-matching algorithm. If we have a perfect $b$-matching, then reachability from $C_a$ to $C_b$ is possible, and is impossible otherwise.

We show that a perfect $b$-matching exists if and only if we can reach the empty configuration. In the forward direction, if there exists a perfect b-matching for the graph corresponding to configuration $C_d$, there exists an assignment $f:e \in E \rightarrow \mathcal{N}$ such that $\sum_{e \in \delta(v)}f(e) = b(v)$ for all $v \in V$. Since all edges $e \in \delta(v)$ correspond to a (2, 0) rule that deletes the species corresponding with $v$, we apply the rule corresponding to $e$ $f(e)$ times. Since $\sum_{e \in \delta(v)}f(e) = b(v)$ where $b(v)$ is the count of the respective species in the configuration, we are deleting all instances of the species using the rules corresponding with $e \in \delta(v)$. Since this is true for all $v \in V$ in a perfect $b$-matching, if there exists a perfect $b$-matching, the empty configuration is reachable. 

For the reverse direction, if we can reach the empty configuration, there exists a sequence of (2, 0) void rules such that we delete all species.  We can construct a perfect $b$-matching using any sequence of (2, 0) void rules that deletes all species. To delete all of a species using (2, 0) void rules the number of (2, 0) void rules applied involving a species is equal to the count of a species. This is because each application of a (2, 0) void rule involving a species deletes the species. We can then see if we can assign $f(e)$ equal to the number of times the corresponding (2, 0) void rule appears in our sequence, this leads to $\sum_{e \in \delta(v)}f(e) = b(v_i)$ for all $v_i \in V$.
Thus, if we can reach the empty configuration using a sequence of (2, 0) void rules, we have a perfect $b$-matching.
\end{proof}

\section{Conclusion}\label{sec:conc}
With the complexity of the general reachability problem solved recently, this paper resolves several restricted cases of the problem. We prove hardness for several open problems or improve the known results. These include showing reachability in Population Protocols with non-increasing volume CRNs are PSPACE-complete with rules of size two, which was originally shown in \cite{Esparza2019}, but we improve this by showing it is still true with only 2-source, 2-consuming CRNs. We prove that feed-forward systems are NP-complete and give a polynomial algorithm if it is single-source or single-consuming without void/autogenesis rules, and showing how void and autogenesis rules affect the complexity of feed-forward systems. Additionally, we provide several other results related to these restrictions.


\para{Related Problems.} While we give many results, this is by no means the end of the investigation into the computational complexity of restricted Chemical Reaction Networks, as this work can be extended in multiple ways:
In general CRNs, it is well known that reachability and production have different complexity.  But in the restricted versions of reachability studied here we have found the problems do not differ in complexity.  At what point, or for what restrictions, do the differences in complexity first merge?  Another variant of reachability is \emph{universal reachability} in which we ask if all system trajectories reach a target configuration.  How does the complexity of universal reachability compare to reachability?  What about the version of production where we want to produce $k$ copies of a given species rather than just a single copy?

\para{Reachability.} For a complete characterization of reachability for all parameters, here we note some open problems and future directions of investigation.

\vspace{-.1cm}
\begin{itemize}
	\item Reachability for 1-consuming \emph{and} 1-source CRNs with the feed-forward property has membership in P. Is the feed-forward restriction required for this result, or does it hold for a 1-consuming and 1-source proper CRN as well?

	\item We show the problem of reachability is PSPACE-complete with rules of size $(2,2)$ (as previously shown in \cite{Esparza2019}).  Is reachability also hard with smaller rule sizes? Can we achieve the same result with $(2,1)$ rules and a binary encoded volume? 
 
    \item What is the smallest catalytic (2,2) system that is PSPACE-complete?
	
	\item Production with $(1,1)$ rules is NL-complete. However, we do not know if reachability is easy as well. Is there a polynomial time algorithm to decide unimolecular reactions?
	
	\item With a non-monotone volume, we no longer have membership in PSPACE- we only have an Ackermann upper bound. With constant rule size, is reachability still Ackermann-hard?
	

\end{itemize}

\bibliographystyle{plain}
\bibliography{crn}

\end{document}